\documentclass[12pt]{iopart}
\usepackage{iopams}  
\usepackage{epsfig}
\begin{document}
\title{Drying layer near a weakly attractive surface}
\author{Alla Oleinikova , Ivan Brovchenko and Alfons Geiger}  
\address{Physical Chemistry, Dortmund University, 44221 Dortmund, Germany }
\eads{\mailto{alla@heineken.chemie.uni-dortmund.de},\mailto{brov@heineken.chemie.uni-dortmund.de}
\mailto{alfons.geiger@udo.edu}}
\begin{abstract}
Depletion of the liquid density near a solid surface with a weak long-range
fluid-surface interaction  was studied by computer simulations of the
liquid-vapor coexistence of a LJ fluid confined in slitlike pores. 
In a wide temperature range the liquid density decreases towards the
surface without the formation of a {\it  vapor} layer between the liquid and the solid
surface. This evidences the absence of a drying transition up to the
liquid-vapor critical point. Two contributions to the excess desorption {\it $\Gamma_{tot}$}  
were found. The first one {\it $\Gamma_{\xi}$} $\sim$ $\rho_{bulk}$ $\xi$
exists at any temperature and diverges as the bulk correlation length $\xi$ when 
approaching the liquid-vapor critical temperature {\it {T$_c$}}. The second
contribution {\it $\Gamma_L$} $\sim$  $\rho_{bulk}$ {\it L$_0$} originates 
from a microscopic {\it drying layer} near the solid
boundary. At high temperatures the thickness {\it L$_0$} of the drying layer increases
in accordance with the power law {\it L$_0$ $\sim$ - ln (1-T/T$_c$)}, indicating a drying transition at {\it {T$_c$}}. The {\it drying layer} can be suppressed by strengthening the
fluid-surface interaction, by increasing the fluid-surface interaction range
or by decreasing the pore size.
\end{abstract}
\maketitle

\section{Introduction}Knowledge of the liquid density profiles  near weakly
 attractive surfaces is necessarry  for the understanding of various
phenomena, such as hydrophobic attraction between extended surfaces in water
\cite{Christ}, slipping flow of liquids near a weakly attractive surface
\cite{Vinogr294}, conformational stability  of biomolecules in aqueous
solutions, etc. Experimental studies evidence a depletion of the liquid
density near weakly attractive substrates
\cite{Chan,Tolan,Fin1,Jensen,Pramana,Grunze}. However, the shape of
the liquid density profiles could not be derived unambigously from
experimental data.
\par In the vast majority of practically important situations, the liquid is close
to equilibrium with the vapor, i.e. close to the liquid-vapor coexistence. 
At the liquid-vapor coexistence curve the vapor (liquid) phase undergoes a wetting (drying) transition near the solid boundary at some temperature {\it T$_{w}$} ({\it T$_{d}$})
 \cite{Cahn,NF82,Pandit,Dietrichrev}. Such a surface transition appears as the
 formation of a macroscopically thick liquid layer in the vapor phase near a strongly
 attractive surface (wetting transition) or as a vapor layer in the liquid phase near
 a weakly attractive surface (drying transition). When crossing {\it T$_{w}$} ({\it T$_{d}$})
 along the liquid-vapor coexistence curve, the liquid (vapor) layer could
 appear continuously (second-order or critical wetting/drying) or discontinuously (first order wetting/drying).
\par Wetting transitions were extensively studied theoretically, experimentally and by
computer simulations for various
 systems. The temperature of the wetting transition and its order
were found to be strongly sensitive to the details of the fluid-fluid and 
fluid-surface interaction \cite{Dietrichrev}. 
In particular, in model systems  with short-range fluid-surface
potential, the wetting transition can be of first or second order, depending on
the strength of the fluid-surface potential \cite{NF82,Binderwet,Swol89}.  The
long-range fluid-surface interaction can change drastically the character and
temperature of the wetting transition. Critical wetting transitions become
first-order due to long-range fluid-surface interactions
\cite{Ebner}. The interplay between the short-range and long-range fluid-surface interaction potentials can
suppress critical wetting up to the liquid-vapor critical temperature
\cite{Indekeu} and produce a sequential (multiple) wetting transition \cite{Bonn2}.
\par   
A drying transition which is accompanied by the formation of a liquid-vapor
interface at some distance from the substrate was observed for short-range
fluid-wall interactions (hard-wall \cite{Hend,Evans2003,Evans2005}, square-well
\cite{Swol89,Swol91} and truncated LJ interactions \cite{Nij}) in computer
simulations and in density functional studies. Contradictory reports
concerning the order of the drying transition in systems with short-range
fluid-wall interaction were discussed in \cite{Hen92}. A long-range fluid-surface
potential (e.g. via van der Waals forces) suppresses a drying transition at
subcritical temperatures, which then can occur at the bulk critical temperature {\it T$_c$} only \cite{Ebner,Indekeu,Bruin}. Since a long-range interaction between
fluids and solids is  unavoidable in real systems, the formation of a macroscopic
vapor layer between the liquid and the surface is impossible. Indeed, a drying
transition was never observed experimentally \cite{Chan,NeCs}.
\par 
Although a macroscopic vapor layer can not appear near a surface with weak long-range
attractive potential, an effect distantly related to a drying transition could be
expected in the liquid phase below {\it T$_c$} as the appearance of a \textit{drying layer}
near the surface (which is not a macroscopic vapor layer and could be considered as "embryo of drying" \cite{Dietrichpriv}). This drying layer can noticeably influence the
liquid density profile near a solid surface.
\par 
Due to the necessary boundary conditions the behavior of the fluid density near a planar solid substrate 
can only be studied by computer simulations in a pore geometry. Most of the computer simulation studies of wetting and drying transitions were performed in \textit{NVT} ensembles, where slitlike pores with symmetrical or asymmetrical walls were incompletely filled (see, for example Refs.\cite{Nij,Bruin}. In such simulations the average density of the confined fluid is deeply inside the two-phase region and a correct reproduction of the liquid-vapor coexistence  is questionable. Even when the liquid-vapor interface in such a pore is well established, it is formed parallel to the wall and, therefore, the wetting phase is represented in such a system as a wetting layer only. In our computer simulation studies of surface transitions we use another approach: simulations of the liquid-vapor coexistence curve of fluids, confined in pores of various sizes, with subsequent extrapolation of the results to semi-infinite systems. 
\par
The chemical potential of the liquid-vapor phase transition of the confined fluid is shifted with respect to the bulk. This suppresses the formation of a wetting or liquid layer along the liquid-vapor coexistence of the confined fluid below and above the temperature of the wetting transition, respectively \cite{Dietrichrev}. To study the possible appearance of a drying (or vapor) layer in the saturated liquid near a weakly attractive surface, we have investigated the temperature evolution of the density profiles along the pore coexistence curve. The data obtained for fluids in pores of various sizes and with different fluid-surface potentials, are compared with the available theoretical predictions and experimental observations.
\par 
 A LJ fluid was confined in pores of width \textit{H} = 40 $\sigma$ ($\sigma$ is the molecular diameter) with weakly attractive walls, which interact with the molecules of the fluid via a long-range potential.  The well depth of the fluid-surface potential was about 70$\%$ of the well depth of the fluid-fluid potential. This is close to the interaction between neon atoms and a cesium substrate, the weakest known physisorption system, recently studied experimentally and by simulations \cite{Chan,NeCs}. In our recent paper \cite{lj1}, the liquid-vapor coexistence curve of a LJ fluid was simulated in a smaller pore with {\it H} = 12 $\sigma$) with the same fluid-surface porential. A depletion of the liquid density near the pore wall was observed in a wide temperature range along the pore coexistence curve without any trend towards the formation of a vapor or drying layer. It was found, that the fluid near the wall follows the universal power laws of the surface critical behaviour of Ising systems \cite{Binderrev,Diehl}. In particular, a universal behavior of the local order parameter profiles, defined as the difference between the local densities $\rho_l$($\Delta${\it z},$\tau$) and $\rho_v$($\Delta${\it z},$\tau$) of the coexisting liquid and vapor
 phases at some distance $\Delta${\it z} from the pore walls, was observed in a wide temperature range \cite{lj1,water2,handbook}. 
The intrusion of the surface critical behavior into the bulk fluid is governed by the bulk correlation length $\xi_-$, which diverges when approaching the critical point:
\begin{eqnarray} 
\label{ksi}
\xi_-(\tau) = \xi_0 \tau^{ -\nu}  
\end{eqnarray}
where $\tau$ = {\it (1 - T/T$_c$)} is the reduced deviation of the temperature {\it T}  from the bulk critical temperature {\it T$_c$}, $\nu$ = 0.63 \cite{nu} is the universal critical exponent and $\xi_0$ is the system dependent amplitude. 
\par 
In view of the high sensitivity of the surface phase transitions to small changes of the chemical potential, the observation of a vapor or drying layer in a pore with {\it H} = 12 $\sigma$ was, probably, prevented by the shift of the liquid-vapor phase transition in
such a rather small pore. To reduce the influence of confinement and to
promote the observation of a vapor or drying layer, the pore width should be as large as
possible. In the present paper we approach the bulk coexistence by performing simulations in a very large pore of width {\it H} = 40 $\sigma$. To extrapolate the results to a semi-infinite system the liquid-vapor coexistence of the LJ fluid in several pores of intermediate sizes between {\it H} = 12 and 40 $\sigma$ was simulated.   
\begin{figure} 
\begin{center}
\includegraphics [width=9cm]{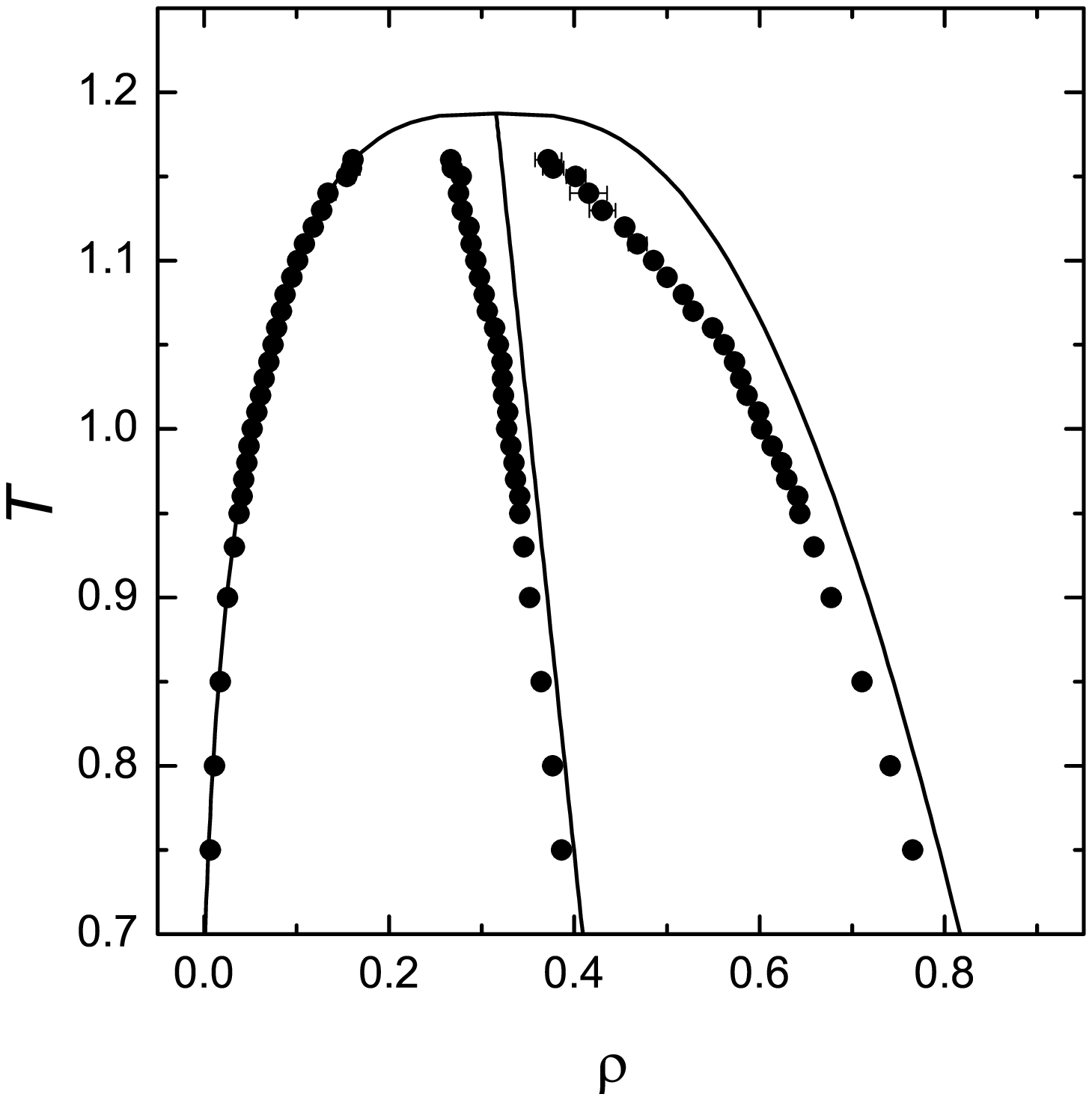}
\caption{Coexistence curve and diameter of the LJ fluid confined in the slitlike pore
of width {\it H} = 40 $\sigma$ (circles). The bulk coexistence curve and
diameter of the LJ fluid \cite{lj1} are shown by solid lines.}
\end{center}
\end{figure}
\section{Method}
Coexistence curves of a confined LJ fluid were determined using Monte Carlo simulations in the Gibbs ensemble (GEMC) \cite{GE}.  GEMC simulations allow to achieve direct equilibration of the two coexisting phases, which are simulated at a given temperature simultaneously
in two simulation cells. The details of the simulations, parameters of the model fluid, as
well as its bulk coexistence curve are given in our previous paper \cite{lj1}. 
To study the behavior of the liquid near the surface, the LJ fluid was confined in a
slitlike pore with structureless walls. Each wall interacts with the particles of
the fluid via the long-range potential of a single plane of LJ
molecules:
\begin{eqnarray}
\label{poten1}
U_w(z) = 4\epsilon\textsl{ f }\left[0.4 \left(\sigma/z\right)^{10}-\left(\sigma/z\right)^4\right],   
\end{eqnarray}
where {\it{z}} measures the distance to the wall and the parameter $\textsl{f}$ determines the strength of the fluid-wall interaction relatively to fluid-fluid interaction and $\sigma$ is the diameter of the LJ particles. No truncation was applied to {\it{U$_w$(z)}}. In the present
paper we report the results obtained for a pore of width $\textit{H}$ = 40 $\sigma$ and {\it{U$_{w1}$(z)}} =  {\it{U$_w$(z, f = 0.3)}}, which corresponds to a weakly attractive (solvophobic) surface with a well-depth of the fluid-wall interaction of about 70$\%$ of the well-depth of the fluid-fluid interaction. To study the effect of the pore size on the liquid density profile and the appearance of a drying layer, we also simulated the liquid-vapor coexistence and the liquid density profiles at {\it T} = 1.10 in pores of width $\textit{H}$ = 16, 22 and 26 $\sigma$ with the same fluid-wall 
interaction {\it{U$_w$(z)}}. 
\begin{figure}
\begin{center}
\includegraphics [width=9cm]{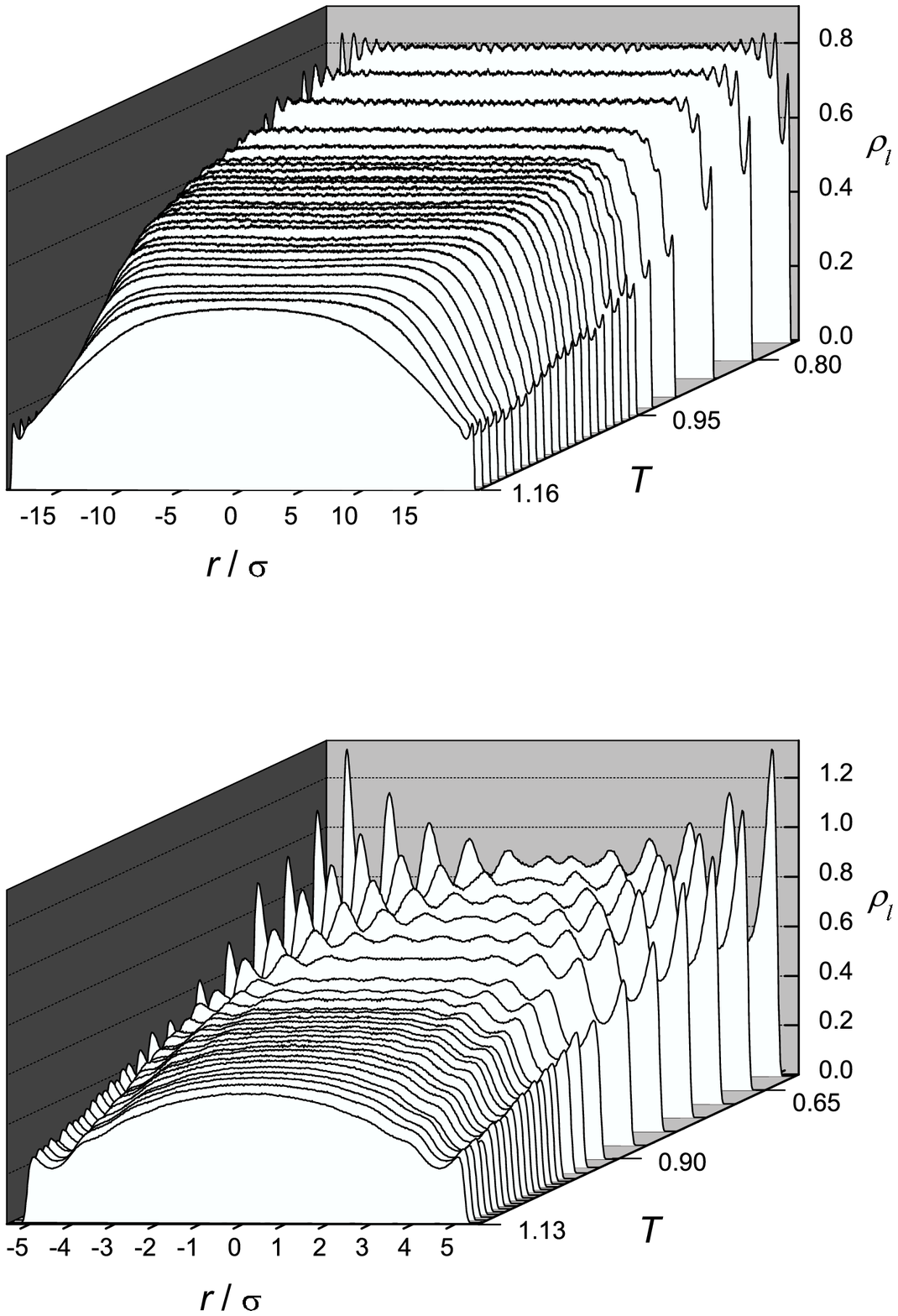}
\caption{Temperature evolution of the liquid density profile along the pore coexistence curves of the LJ fluid confined in a slitlike pore with {\it H} = 40 $\sigma$ (upper panel) and {\it H} = 12 $\sigma$ (lower panel).}
\end{center}
\end{figure}  
\par
To explore the effect of the strength of the fluid-wall potential on the liquid density profile we
also simulated the liquid-vapor coexistence in a pore with slightly stronger fluid-surface interaction {\it{U$_{w2}$(z)}} =  {\it{U$_w$(z, f = 0.4)}}. Additionally, the effect of the range of the fluid-wall interaction was explored by simulation of a system with a slower decay of the fluid-wall potential {\it{U$_{w3}$(z)}}, which follows from an integration of the individual interactions over the half-space of LJ molecules and can be described by the equation:
\begin{eqnarray}
\label{poten2}
U_{w3}(z) = 4\epsilon {\it f^*}\left[\left(\sigma^*/z\right)^{9}-
\left(\sigma^*/z\right)^3\right],   
\end{eqnarray} 
The parameters of the potential {\it{U$_{w3}$(z)}} were adjusted to get equal well depths of the potentials {\it{U$_{w3}$(z)}} and {\it{U$_{w1}$(z)}}. This was achieved using $\sigma^*$ = 0.8328 $\sigma$ and {\it  f$^*$} = 0.468 in equation (\ref{poten2}).
\par 
The density $\rho$ used in the present paper is the reduced number density (scaled by $\sigma^3$), while {\it{T}} is the reduced temperature (scaled by $\epsilon$/$\textit{k}_{B}$, where $\textit{k}_{B}$ is a Boltzmann`s constant). The average density of the LJ fluid confined in a pore was calculated taking
into account the volume accessible to the fluid molecules. The fluid-wall interaction is equal to zero at the distance 0.86 $\sigma$ from the pore wall and so in an operational approach this interval could be divided equally between the volumes of the fluid and the solid. As a  result, the volume accessible to the fluid molecules is the total pore volume
 $\textit{L$^2$H}$ reduced by a factor 1.022 for the pore of width {\it H} = 40  $\sigma$. Accordingly, as the distance to the pore wall \textit{$\Delta$z} we used the distance to the parallel plane 0.5 $\sigma$ inside the pore.
\par 
The density profiles of the liquid were obtained by Monte Carlo (MC) simulations in the {\it NVT}
 ensemble, using the average densities of the liquid phase obtained in the GEMC simulations. At high temperatures, the strong density gradient near the pore wall makes the determination of reliable density profiles in such large pores very time consuming. This problem was overcome by using two kinds of moves in MC simulations in the {\it NVT} ensemble. The first kind of moves is a standard MC move with a maximal displacement of molecule, which provides an acceptance probability of about 50 $\%$. The second kind of move is a long-distance molecular transfer inside the simulation box: an attempt to place randomly chosen fluid molecule into randomly chosen position. This move is similar to the one used in GEMC simulations for molecular transfers between the two simulation boxes. Such long-distance molecular transfers essentially improve the sampling of density profiles, which show a strong gradient normal to the pore wall.    
\section{Results}
\begin{figure}
\begin{center} 
\includegraphics [width=9cm]{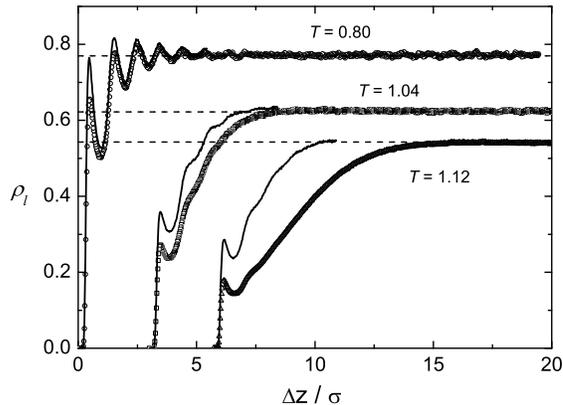}
\caption{The density profiles of the coexisting liquid phase of the LJ fluids confined in the slitlike pores with {\it H} = 40 $\sigma$ (symbols) and 12 $\sigma$ (solid lines). The profiles
for {\it T} = 1.04 and {\it T} = 1.12 are shifted to larger \textit{$\Delta$z} by 3 and 6
$\sigma$, respectively. }
\end{center}
\end{figure} 
The obtained densities of the coexisting vapor ($\rho_v$) and liquid ($\rho_l$) phases and the
diameter (($\rho_l$ + $\rho_v$)/2) of the coexistence curve of the LJ fluid in the
slitlike pore with weakly attractive walls of width {\it H} = 40 $\sigma$ are shown in figure 1. The critical temperature of the pore coexistence curve is depressed due to the effect of the confinement. It is estimated as {\it T} = 1.165 $\pm$ 0.005, that is only about 0.02 below the bulk critical temperature {\it T$_c$} = 1.1876 \cite{Wilding,lj1}. This indicates that in this case the pore coexistence curve is rather close to the coexistence curve of the bulk fluid. In such a wide pore, we may therefore expect that below the pore critical temperature the properties of the fluid near a wall are only slightly influenced by the presence of the opposite wall.
\begin{figure}
\begin{center} 
\includegraphics [width=7cm]{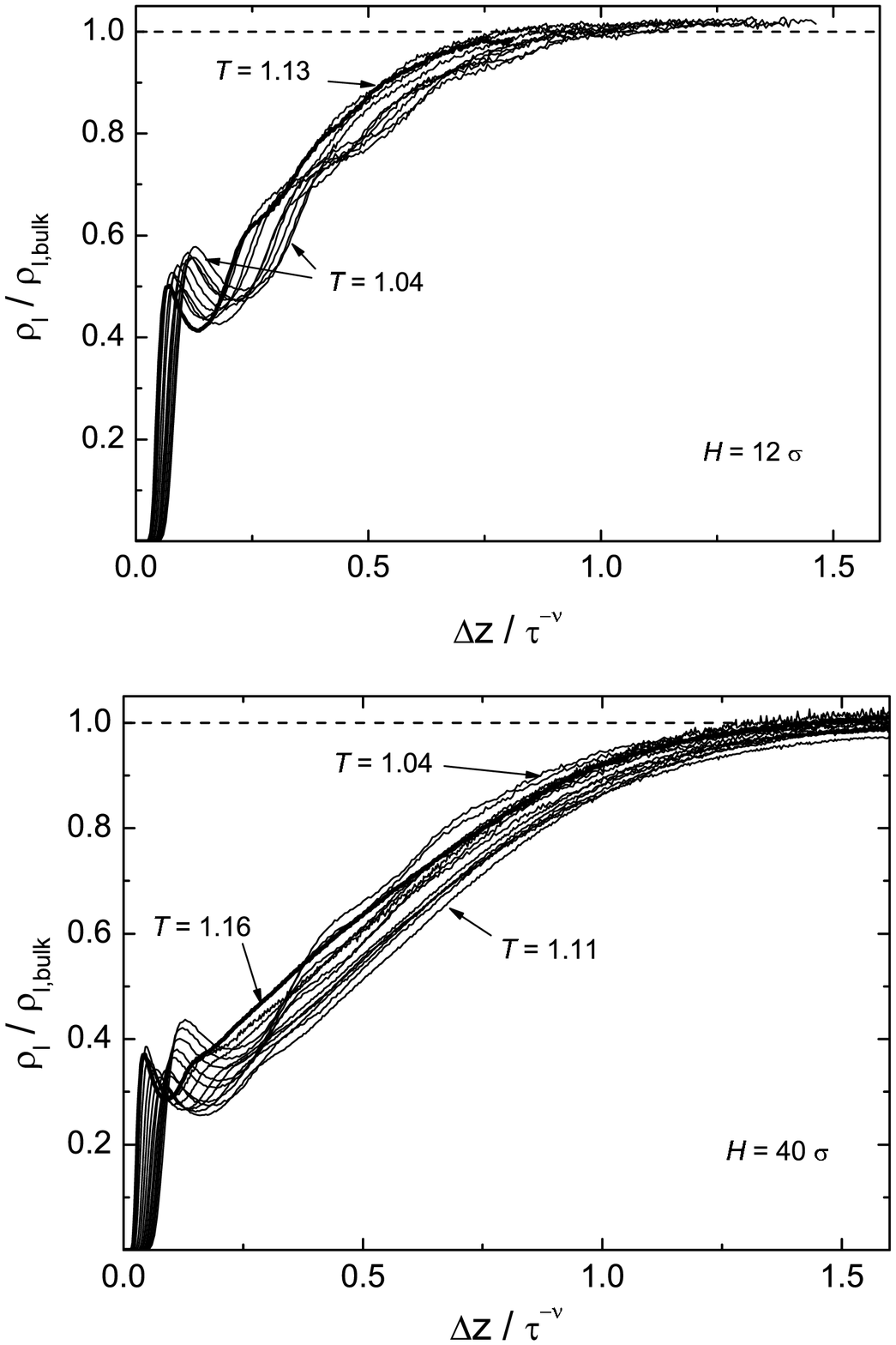}
\caption{Scaling plot for the liquid density profiles in the small pore with
{\it H} = 12 $\sigma$ for {\it T} = 1.04, 1.05, 1.06, 1.07, 1.08,
1.09, 1.10, 1.11, 1.12 and 1.13 (upper panel) and in the large pore with {\it H} = 40
$\sigma$ for {\it T} = 1.04, 1.05, 1.06, 1.07, 1.08,
1.09, 1.10, 1.11, 1.12, 1.13, 1.14, 1.15 , 1.155 and 1.16 (lower panel). The
highest studied temperature for each pore is shown by a thick solid line.}
\end{center}
\end{figure} 
 \par Figure 1 shows that the average density of the liquid phase at the pore coexistence curve is noticeably lower than the density of the bulk LJ fluid at the same temperature. This effect becomes more pronounced at higher temperatures, i.e. at about {\it T} $>$ 1.05 and it originates from the depletion of the liquid density near a weakly attractive wall \cite{lj1,water2,handbook,diff2003,water1}. Indeed, a pronounced decrease of the liquid density near the wall is observed in the large ({\it H} = 40 $\sigma$) as well as in the small ({\it H} = 12 $\sigma$ \cite{lj1}) pore (see figure 2). At high temperatures the liquid density profile is flat in the pore interior (constant  $\rho_l$($\Delta$z,$\tau$) in the central part of the pore) and gradually decreases toward the pore wall, displaying only a single small oscillation near the pore wall (figure 2, upper panel). Upon cooling, the region of density depletion shrinks 
and additional density oscillations develop gradually on the density profiles.  A qualitatively similar behavior of the density profile is seen in the liquid phase in the small pore of width {\it H} = 12 $\sigma$ (figure 2, lower panel). In the small pore, however, the density depletion spreads over the whole pore near the pore critical temperature and the flat part of liquid density profile is not observed. 
\begin{table}[t]
\caption{The values of the fitting parameters $\xi_-$, \textit{L$_0$} and
  $\rho_{l,bulk}$ in equation \ref{tanh} when fitted to the liquid density profiles in the large pore of width {\it H} = 40 $\sigma$. Uncertainties of the fitting parameters correspond to the confidence level 95$\%$. The values of $\rho_{l,bulk}^0$ were obtained by direct GEMC simulations of the liquid-vqpor equilibrium \cite{lj1}. }
\vspace{0.5cm} 
\begin{center}
\begin{tabular}{c|c|c|c|c}
\hline\noalign{\smallskip}
\textit{T} & $\xi_-$ / $\sigma$ & \textit{L$_0$} / $\sigma$ & $\rho_{l,bulk}$
& $\rho_{l,bulk}^0$ \\
 & $\pm$0.01 & $\pm$0.01 & $\pm$0.0005 &  \\
\noalign{\smallskip}\hline\noalign{\smallskip}
0.75 & 0.19 & 0.92 & 0.7935 & 0.793 \\
0.80 & 0.24 & 0.89 & 0.7703 & 0.769 \\
0.85 & 0.35 & 0.83 & 0.7424 & 0.743 \\
0.90 & 0.56 & 0.84 & 0.7161 & 0.717 \\
0.93 & 0.62 & 0.89 & 0.6996 & 0.697 \\
0.95 & 0.69 & 0.99 & 0.6880 & 0.684 \\
0.96 & 0.66 & 0.92 & 0.6826 & 0.678 \\
0.97 & 0.75 & 1.05 & 0.6764 & 0.672 \\
0.98 & 0.67 & 0.95 & 0.6649 & 0.665 \\
0.99 & 0.81 & 1.14 & 0.6634 & 0.659 \\
1.00 & 0.87 & 1.29 & 0.6560 & 0.651 \\
1.01 & 0.82 & 1.17 & 0.6484 & 0.644 \\
1.02 & 0.82 & 1.23 & 0.6362 & 0.637 \\
1.03 & 0.89 & 1.31 & 0.6328 & 0.628 \\
1.04 & 0.87 & 1.31 & 0.6247 & 0.622 \\
1.05 & 0.95 & 1.40 & 0.6164 & 0.613 \\
1.06 & 1.10 & 1.57 & 0.6103 & 0.605 \\
1.07 & 1.16 & 1.73 & 0.5920 & 0.597 \\
1.08 & 1.16 & 1.84 & 0.5830 & 0.590 \\
1.09 & 1.31 & 1.23 & 0.5760 & 0.576 \\
1.10 & 1.47 & 2.26 & 0.5630 & 0.568 \\
1.11 & 1.57 & 2.69 & 0.5550 & 0.558 \\
1.12 & 1.68 & 2.71 & 0.5420 & 0.547 \\
1.13 & 1.95 & 2.87 & 0.5210 & 0.529 \\
1.14 & 2.14 & 3.18 & 0.5140 & 0.524 \\
1.15 & 2.40 & 3.28 & 0.5036 & 0.502 \\
1.155 & 2.91 & 3.69 & 0.5065 & 0.495 \\
1.16 & 3.32 & 3.75 & 0.493 & 0.488 \\
\noalign{\smallskip}\hline
\end{tabular}
\end{center}
\footnotetext[1]{data for the bulk coexistence curve \cite{lj1}}
\end{table} 
\par
A direct comparison of the density profiles of the saturated liquid in the small and large pores is shown in figure 3 for some temperatures. At {\it T} = 0.80 the profiles are practically identical in both pores. The density oscillations are slightly larger in the small pore. At high temperatures, the density depletion is weaker in the small pore and the difference between $\rho_l$($\Delta$z,$\tau$) in the small and the large pore becomes more pronounced with increasing temperature. 
\par 
  It was shown \cite{lj1,handbook} that the local order parameter
profiles $\Delta\rho$($\Delta${\it z},$\tau$) = ($\rho_l$($\Delta${\it
  z},$\tau$) -  $\rho_v$($\Delta${\it z},$\tau$))/2, in the small pore of width {\it H}
= 12 $\sigma$ with weakly attractive walls collapse into a single master curve, 
when the local order parameter is normalized by the bulk order parameter, and the distance to the pore wall is normalized by the bulk correlation length (see figure 9 in
Ref. \cite{lj1}). Such a universal behavior of $\Delta\rho$($\Delta${\it
  z},$\tau$) evidences that the drying transition, which is expected at the bulk
critical temperature, does not influence the liquid density profiles in such a small
pore. In the large pore of width {\it H} = 40 $\sigma$, we do not obtain such 
universal behavior of $\Delta\rho$($\Delta${\it z},$\tau$) when using the same
approach, as described above \cite{lj1}. We have found, that this discrepancy originates
from the quite different temperature dependences of the density profiles of the \textit{liquid} phase in small and large pores. 
\begin{figure}
\begin{center} 
\includegraphics [width=9cm]{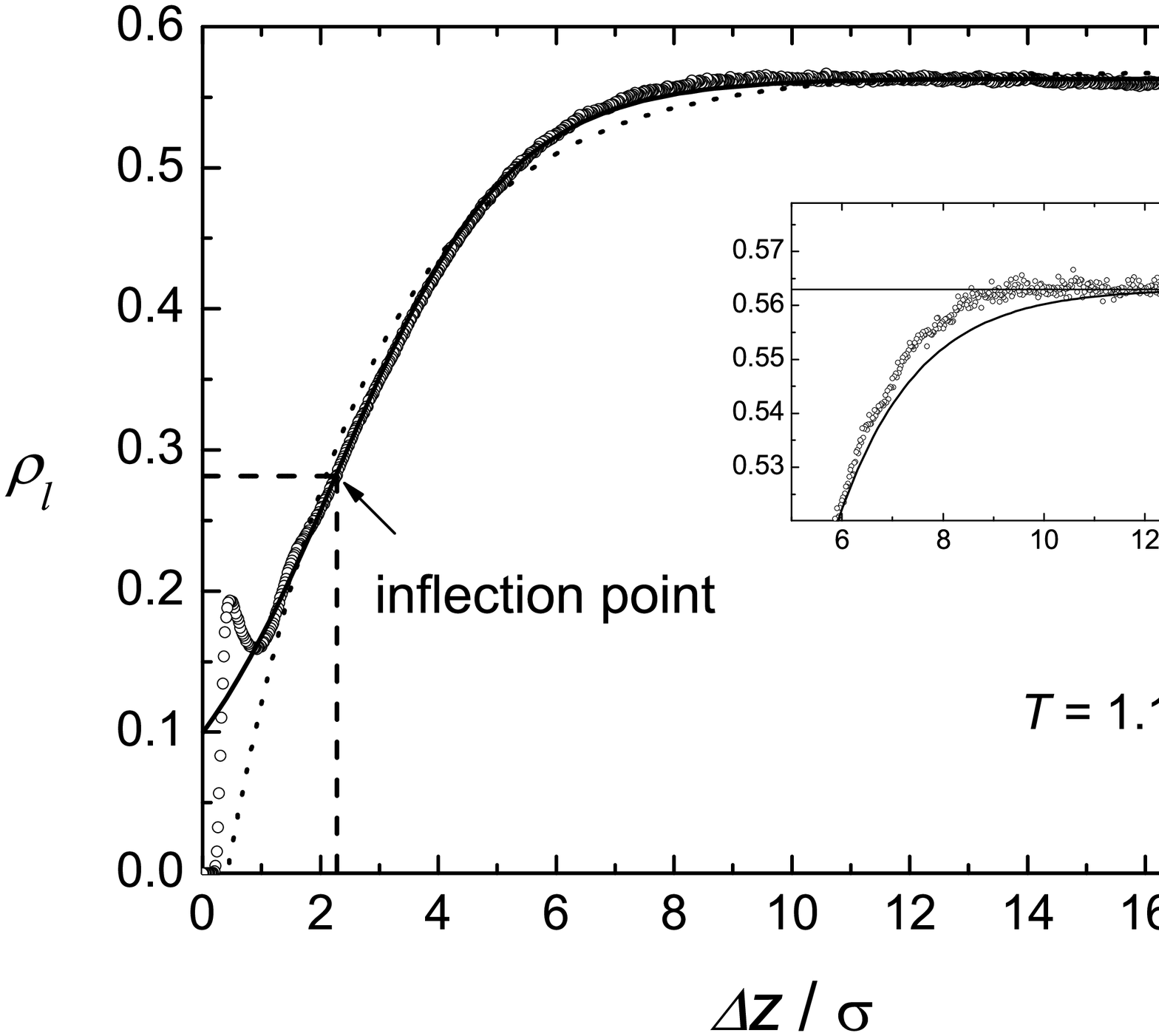}
\caption{Liquid density profile in the large pore with {\it H} = 40 $\sigma$ at {\it T} = 1.10 (circles) and the fit to equation \ref{tanh} with the parameters shown in Table 1 (solid line) and to equation \ref{exp} (dotted line). Coordinates of the inflection points are shown by dashed lines. The inset gives a sectoral magnification. The value $\rho_{l,bulk}$ found from the fit is shown by the horizontal line in the inset.}
\end{center}
\end{figure}
\par 
The liquid density profiles at various temperatures can be compared, using the normalized
density $\rho_l$($\Delta${\it z},$\tau$)/$\rho_{l,bulk}$($\tau$), where $\rho_{l,bulk}$($\tau$) is the liquid density of the bulk LJ fluid (the values, obtained in Ref. \cite{lj1}, are shown in the last column of Table 1). Because the intrusion of the surface perturbation into the
bulk fluid is governed by the bulk correlation length $\xi$, it is reasonable
to display the distance to the pore wall in terms of $\xi$. Taking into account the
universal temperature dependence of $\xi$ along the coexistence curve
(equation (\ref{ksi})), we can use  $\Delta${\it z}/$\tau^{-\nu}$ as a normalized
distance variable. The normalized density profiles of the liquid in the small and the large
pore are compared in figure 4 in the temperature range from {\it T} = 1.04 up 
to the highest temperature of the two-phase coexistence in the pore. 
\begin{figure}
\begin{center} 
\includegraphics [width=8cm]{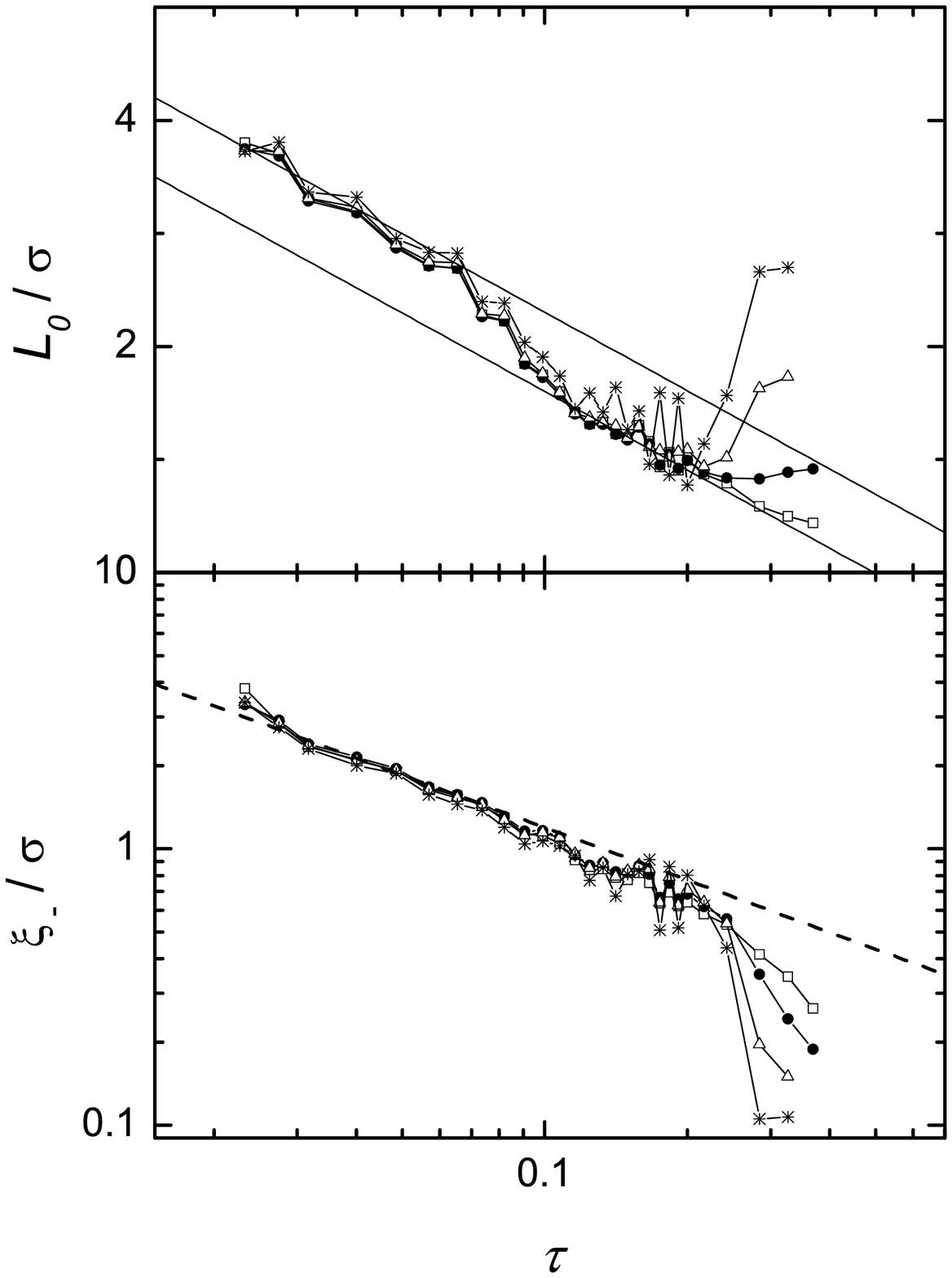}
\caption{Temperature dependence of the thickness of the drying layer {\it L$_0$} and correlation length $\xi_-$, obtained from the fits of equation (\ref{tanh}) to the part of the liquid density profile 
$\rho_l$($\Delta${\textit z},$\tau$), which extends from
the pore center to various distances $\Delta${\textit z$_{cut}$} from the surface:
$\Delta${\textit z$_{cut}$} = 0 $\sigma$ (squares), 1 $\sigma$ (circles), 2 $\sigma$
(triangles) and 3 $\sigma$ (stars). A logarithmic behavior described by
equation (\ref{layer}) with {\it const} =  -0.05 $\sigma$ and  -0.85
$\sigma$ is shown by upper and lower solid lines, respectively. The power law (\ref{ksi})
with $\xi_0$ = 0.28 $\sigma$ is shown by a dashed line.}
\end{center}
\end{figure}
\par
In the small pore, the normalized density profile $\rho_l$($\Delta${\it z},$\tau$)
can be roughly described by the exponential equation:
\begin{eqnarray}
\label{exp}
\frac{\rho_l(\Delta z,\tau)}{\rho_{l,bulk}(\tau)} = \left[1-exp\left(-\frac{\Delta z - 
l_0}{\xi_0\tau^{-\nu}}\right)\right].
\end{eqnarray}
The values of the parameter {\it l$_0$} obtained from the fits do not exceed
a few tenth of $\sigma$, whereas the value of the fitting parameter $\xi_0$ is about 0.3
$\sigma$, close to the value of $\xi_0$, obtained from the master curve of the
order parameter profiles in Ref. \cite{lj1}. In the large pore, the shape of the liquid density profiles is qualitatively different from that in the small pore (figure 4). Moreover, in the large pore the normalized profiles 
$\rho_l$($\Delta$z,$\tau$)/$\rho_{l,bulk}$($\tau$) vary with temperature
non-monotonously with the "lowest" profile at {\it T} = 1.11 (see figure 4). The shape of the
liquid density profiles at any temperature can not be fitted to the exponential equation
(\ref{exp}). 
\par
These peculiarities of the liquid density profiles in the large pore indicate 
the possible presence of some non-universal contribution to the surface behavior. As the increase of
the pore size can be considered as an approach to a semi-infinite system, one may expect that this peculiar behavior is connected with the formation of a drying layer, caused by a drying transition at the bulk critical temperature \cite{Dietrichrev}. If such a drying layer appears in a liquid phase near the surface of a large pore, the liquid density profile should be described by an interface-like equation. The density profile at the intrinsic liquid-vapor interface, derived
from the van der Waals theory, could be described by the following equation \cite{vdW,Widom}:
\begin{eqnarray} 
\label{tanhlv}
\fl \rho_l(\Delta z,\tau) = \frac{\rho_{l,bulk}(\tau)-\rho_{v,bulk}(\tau)}{2}tanh\left(\frac{\Delta z - L_0}{2\xi_-}\right) + \frac{\rho_{l,bulk}(\tau)+\rho_{v,bulk}(\tau)}{2},  
\end{eqnarray}
where {\it L$_0$} is the distance from the liquid-vapor interface to the solid surface.
This intrinsic interfacial profile can be affected by capillary waves, i.e. thermal fluctuations of the interface \cite{cw}. Far from the critical point the bulk fluctuations are small and capillary waves are the dominant factor, which determines the thickness of the interface \cite{Dietrichinter}. Upon rising the temperature the density fluctuations in the bulk
phases become more important, yielding an intrinsic interfacial thickness proportional to the increasing bulk correlation length, as described by equation (\ref{tanhlv}). Since the expected interface is close to the wall, the capillary waves can be suppressed by a long-range fluid-surface interaction. 
\par
We have found, that the liquid density profile in the large pore can be described by equation (\ref{tanhlv}) at low temperatures, when the density of the bulk vapor is close to zero. However, equation (\ref{tanhlv}) fails to fit the liquid density profiles at high temperatures, because the value of the density near the pore wall is even lower than the bulk vapor density at the same temperature. This result is consistent with the theoretical expectations, that a vapor layer never appears near a surface with long-range fluid-surface interaction \cite{Ebner,Indekeu}. So, it is more reasonable to suggest not a \textit{vapor} but a \textit{drying} layer near the surface. Since the density distribution in a drying layer is unknown, we fitted the liquid density profiles to equation (\ref{tanhlv}) with $\rho_{v,bulk}$ as a free parameter $\rho^*$. Such an approach provides a satisfactory description of the simulated profiles apart from the first layer of about $\sigma$ width, where a density oscillation, caused by fluid-surface interaction potential, is noticeable even at supercritical temperatures. The values of $\rho^*$ obtained from the fits were found close to zero for all studied temperatures. An example of such a fit with $\rho^*$ = 0 for {\it T} = 1.10 is shown in figure 5. 
\begin{figure}
\begin{center} 
\includegraphics [width=8cm]{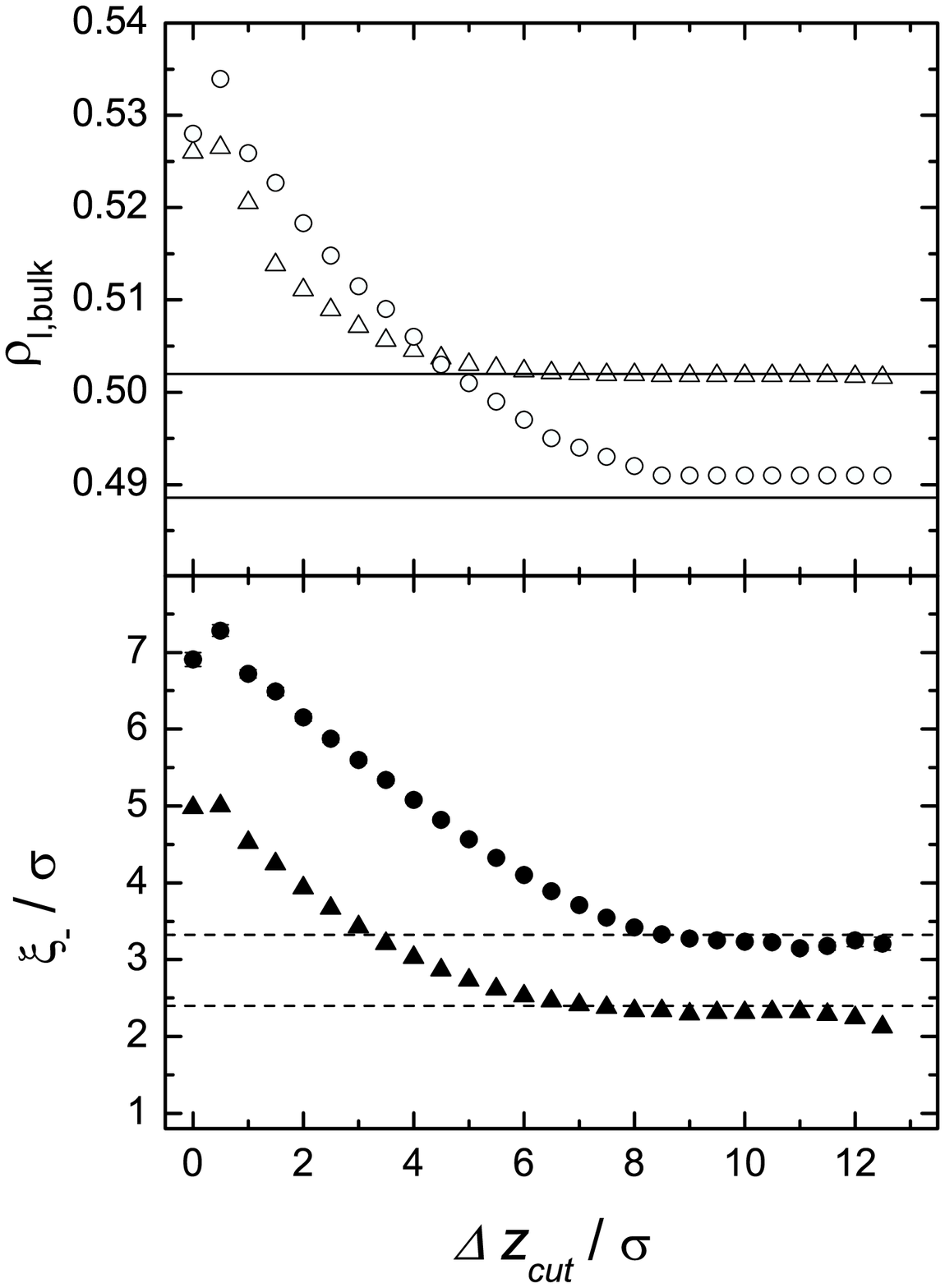}
\caption{Variation of the fitting parameters $\rho_{l,bulk}$ and $\xi_-$ at {\it T} =
1.15 (triangles) and {\it T} = 1.16 (circles), when  equation (\ref{exp}) is fitted to those parts of the liquid density profile $\rho_l$($\Delta${\it z},$\tau$), which extend from
 the pore center to various distances $\Delta${\it z$_{cut}$} from the surface.
The values of the liquid density at the bulk coexistence curve are shown by
 solid lines (upper panel). The values of the bulk correlation length
 $\xi_-$, obtained from the fit of equation (\ref{tanh}) to $\rho_l$($\Delta${\it z},$\tau$) (Table 1) are shown by dashed lines (lower panel). }
\end{center} 
\end{figure}
So, we propose the following equation to describe the liquid density profile
$\rho_l$($\Delta${\it z},$\tau$) near a weakly attractive surface: 
\begin{eqnarray} 
\label{tanh}
\rho_l(\Delta z,\tau) = \frac{\rho_{l,bulk}(\tau)}{2}\left[tanh\left(\frac{\Delta z
      - L_0}{2\xi_-}\right)+1\right].  
\end{eqnarray}
There are three parameters in equation (\ref{tanh}): {\it L$_0$}, the distance
of the inflection point of the such defined liquid-drying layer interface  from the surface (thickness of the drying layer), $\xi_-$, the bulk correlation length at the liquid-vapor coexistence curve and $\rho_{l,bulk}$($\tau$), the density of the bulk liquid. The first density oscillation (which extends over about 1 $\sigma$ near the surface) is caused by the localization of the molecules in the well of the fluid-wall potential. Since it strongly deviates from the smooth density profile outside the first layer, it is reasonable to exclude the first layer from the fits. The values of the parameters obtained from the fits of equation (\ref{tanh}) to the simulated liquid density profiles are shown in Table 1. 
\par
The temperature dependences of the correlation length $\xi_-$ and of the thickness of the drying layer {\it L$_0$}, obtained from these fits, are shown in figure 6 (solid circles). The correlation length shows a temperature dependence, which is rather close to the simple power law described by equation (\ref{ksi}) with an amplitude $\xi_0$ = 0.28 $\sigma$ (figure 6, lower panel, dashed line). Such an amplitude of the correlation length agrees well with the value $\xi_0$ = 0.3
$\sigma$, obtained previously from the master curve of the order parameter of a LJ fluid
in a narrow pore \cite{lj1}. 
\par 
Another estimate of the amplitude of the correlation length can be obtained based on the exponential decay of the perturbation caused by the surface: far from the surface the deviation of the local density from the bulk value is governed by the bulk correlation length
and should follow equation (\ref{exp}). Thus, the progressive exclusion of an increasing part of the density profile near the surface from the fit, should show a convergence of the fitting parameters, including $\xi_-$. The values of $\xi_-$ obtained from fits of equation (\ref{exp}) to liquid density profiles, which extend from the pore center to various distances from the surface {\it $\Delta$z$_{cut}$} are shown in figure 7 (lower panel) for two temperatures. A clear convergence of the fitted values $\xi_-$ is observed, when the fitted part of the density profile does not approach the pore wall closer than 6 $\sigma$ at {\it T} = 1.15 and 8 $\sigma$ at {\it T} = 1.16. These converged values of $\xi_-$ are remarkably close to the values of $\xi_-$ obtained from the fit of the data to equation (\ref{tanh}), shown by dashed lines (figure 7). 
\par
The values $\rho_{l,bulk}$ obtained from fits of $\rho_l$($\Delta${\textit z},$\tau$) to the interfacial equation (\ref{tanh}) (Table 1) are almost equal to the liquid densities in the pore center, obtained by fits of the data near the pore center to the exponential equation (\ref{exp}) (see solid lines in figure 7). The values $\rho_{l,bulk}$ obtained from these fits agree very well with the bulk liquid densities obtained by direct Gibbs ensimble MC simulations of the liquid-vapor equilibrium \cite{lj1}, which are also presented in Table 1. 
\par 
Next, we test the sensitivity of the fitting parameters in the interfacial equation (\ref{tanh}) from the portion of the liquid density profile used for the fits. For this purposes we fitted equation (\ref{tanh}) to the part of $\rho_l$($\Delta${\it z},$\tau$), which extends from the pore center to  some distances {\it $\Delta$z$_{cut}$} from the surface. The fitting parameter $\rho_{l,bulk}$ was found practically independent on the choice of the fitting interval. The values of the correlation length $\xi_-$, obtained from the fits, depend on the choice of $\Delta$z$_{cut}$ (from 0 to 3 $\sigma$) only at {\it T} $\leq$ 0.90 (see figure 6, lower panel) due to the strong density oscillations near the surface at low temperatures. At {\it T} $>$ 0.90 the fitting results for the correlation length $\xi_-$ are not sensitive to {\it $\Delta$z$_{cut}$}. 
\par
The change of the thickness of the drying layer {\it L$_0$} with temperature is shown in the upper panel of figure 6. At the lowest temperatures ({\it T} $\leq$ 0.90) the fitted value of {\it L$_0$} is determined by the choice of $\Delta$z$_{cut}$, namely {\it L$_0$} $\approx$ $\Delta$z$_{cut}$. This correlation is clearly imposed by the strong density oscillations near the surface. Thus, if the thickness of a drying layer does not exceed 1 $\sigma$, meaningful value of {\it L$_0$} can not be determined from the fits. At higher temperatures, {\it L$_0$} does not depend noticeably on the choice of $\Delta$z$_{cut}$. Two temperature regions could be distinguished in the temperature dependence of {\it L$_0$} at {\it T} $>$ 0.90. In the temperature interval from {\it T} = 1.11 to 1.16 the thickness of a drying layer {\it L$_0$} clearly follows the logarithmic dependence:  
\begin{eqnarray}
\label{layer}
L_0 = ln\left(\tau^{-1}\right) + const
\end{eqnarray}
with {\it const} $\approx$ -0.05 $\sigma$ (see Fig.6, upper panel, upper solid line). A quite similar logarithmic temperature dependence of {\it L$_0$} is observed in the low temperature interval from {\it T} = 0.85 to 1.06, but with another value {\it const} $\approx$ -0.85 $\sigma$.  There is a crossover between two kinds of logarithmic behavior in the temperature interval from {\it T} = 1.07 to 1.10. Note, that the observed logarithmic temperature dependence of {\it L$_0$} is similar to that predicted for the divergence of the thickness of the wetting layer in the case of a critical wetting transition \cite{Dietrichrev}. 
\par 
So, equation \ref{tanh} allows the description of the liquid density profiles in a wide temperature range using two properties of the bulk fluid ($\rho_{l,bulk}$ and $\xi_-$) and the thickness of the drying layer {\it L$_0$}. The degree of the universality of the liquid density profiles can be illustrated by a master curve, which is essentially improved in comparison with figure 4, where the presence of a drying layer was neglected. Now, we introduce a normalized length scale, which is measured from the liquid-drying layer interface , as ($\Delta$z - {\it L$_0$})/$\xi_-$. Using the fitting parameters {\it L$_0$} and $\rho_{l,bulk}$ from Table 1, the liquid density profiles collapse on a single master curve in a wide temperature range 1.04 $<$ {\it T} $<$ {\it T$_c$} (see figure 8). The dashed line in figure 8 represents the interfacial equation (\ref{tanh}) in normalized coordinates. This means, that equation (\ref{tanh}) allows a perfect description of the liquid density profiles in the whole temperature range, where the drying layer can be clearly detected. In the temperature range above {\it T} = 1.11, where the thickness of
the drying layer follow equation (\ref{layer}) (see figure 6), $\rho_l$($\Delta${\it z},$\tau$) can be predicted if the bulk density, the bulk correlation length and the {\it const} in equation (\ref{layer}) is known. Note, that the latter value can be estimated from a single density profile.  
\begin{figure}
\begin{center} 
\includegraphics [width=9cm]{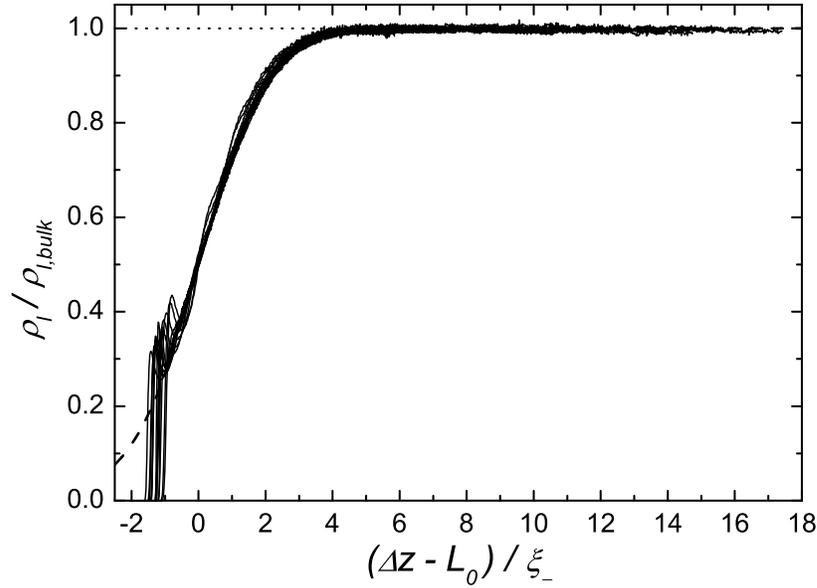}
\caption{Scaling plot of the liquid density profiles $\rho_l$($\Delta$
  z,$\tau$) in the large pore {\it H} = 40 $\sigma$ for the same temperatures as in
  figure 4. The parameters $\rho_{l,bulk}$ and {\it L$_0$} were obtained from
  fits of equation (\ref{layer}) to density profiles $\rho_l$($\Delta$ z,$\tau$), which were
  cut at the distance 1 $\sigma$ from the surface (Table 1). For {\it T} = 1.16 the
  value $\rho_{l,bulk}$ from the bulk coexistence curve was used. }
\end{center}
\end{figure}
\begin{figure}
\begin{center} 
\includegraphics [width=8cm]{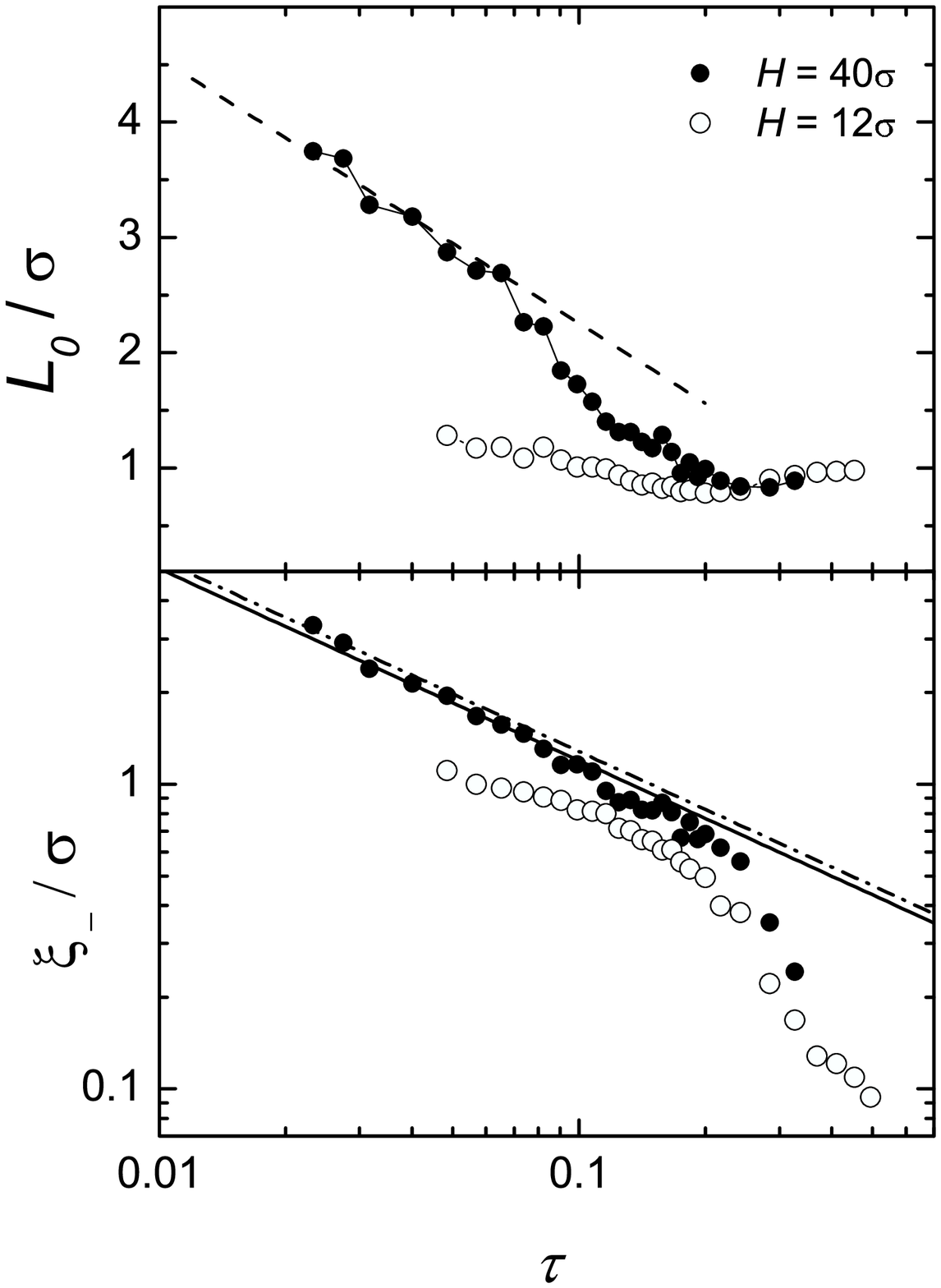}
\caption{Temperature dependence of the thickness of the drying layer
{\it L$_0$} and of the correlation length $\xi_-$, obtained from fits
of the liquid density profiles $\rho_l$($\Delta${\it z},$\tau$) cut at the distance 1 $\sigma$ from the surface for large (solid circles) and small (open circles) pores. Solid and dot-dashed
lines represents equation (\ref{ksi}) with $\xi_0$ = 0.28 $\sigma$ (this
paper) and $\xi_0$ = 0.30 $\sigma$ \cite{lj1}, respectively. Dashed line
corresponds to equation (\ref{layer}) with \textit{const} = -0.05 $\sigma$.}
\end{center}
\end{figure}
\par
 We also test the possibility to describe the density profiles of the liquid in the small pore with {\it H} = 12 $\sigma$ by the interfacial equation (\ref{tanh}). The obtained fitting values of {\it L$_0$} are about 1 $\sigma$ in the whole temperature range (figure 9). The fitting values of {\it L$_0$} in the small and large pores coincide at low temperatures ({\it T} $<$ 0.90), where they are strongly influenced by the density oscillations and, therefore, are meaningless. As far as a thickness of the drying layer below 1 $\sigma$ can not be detected from the fits, the real thickness of the drying layer can be essentially lower. Thus, the fitting values of {\it L$_0$} for the small pore with {\it H} = 12 $\sigma$, shown in figure 9, can not be considered as an indication of a drying layer in this pore even near the pore critical temperature.  Obviously, the fluid layer near the surface, highly localized in the well of a fluid-wall potential, can not be considered as a drying layer. This observation agrees with a previous analysis of the local order parameter in the same pore \cite{lj1}, which indicated the absence of a drying layer. 
\par
The values of $\xi_-$ found from the fits of $\rho_l$($\Delta${\it z},$\tau$)
in the small pore with equation (\ref{tanh}) are noticeably smaller than $\xi_-$ obtained from the fits of the order parameter in the same pore \cite{lj1} and $\xi_-$ in the large pore (figure
9). So, the liquid density profiles in the small pore can not be described correctly by equation (\ref{tanh}): the effective thickness of the drying layer is ultimately overestimated in such a pore, resulting consequently in an underestimation of the value of the correlation length. 
\begin{figure}
\begin{center} 
\includegraphics [width=9cm]{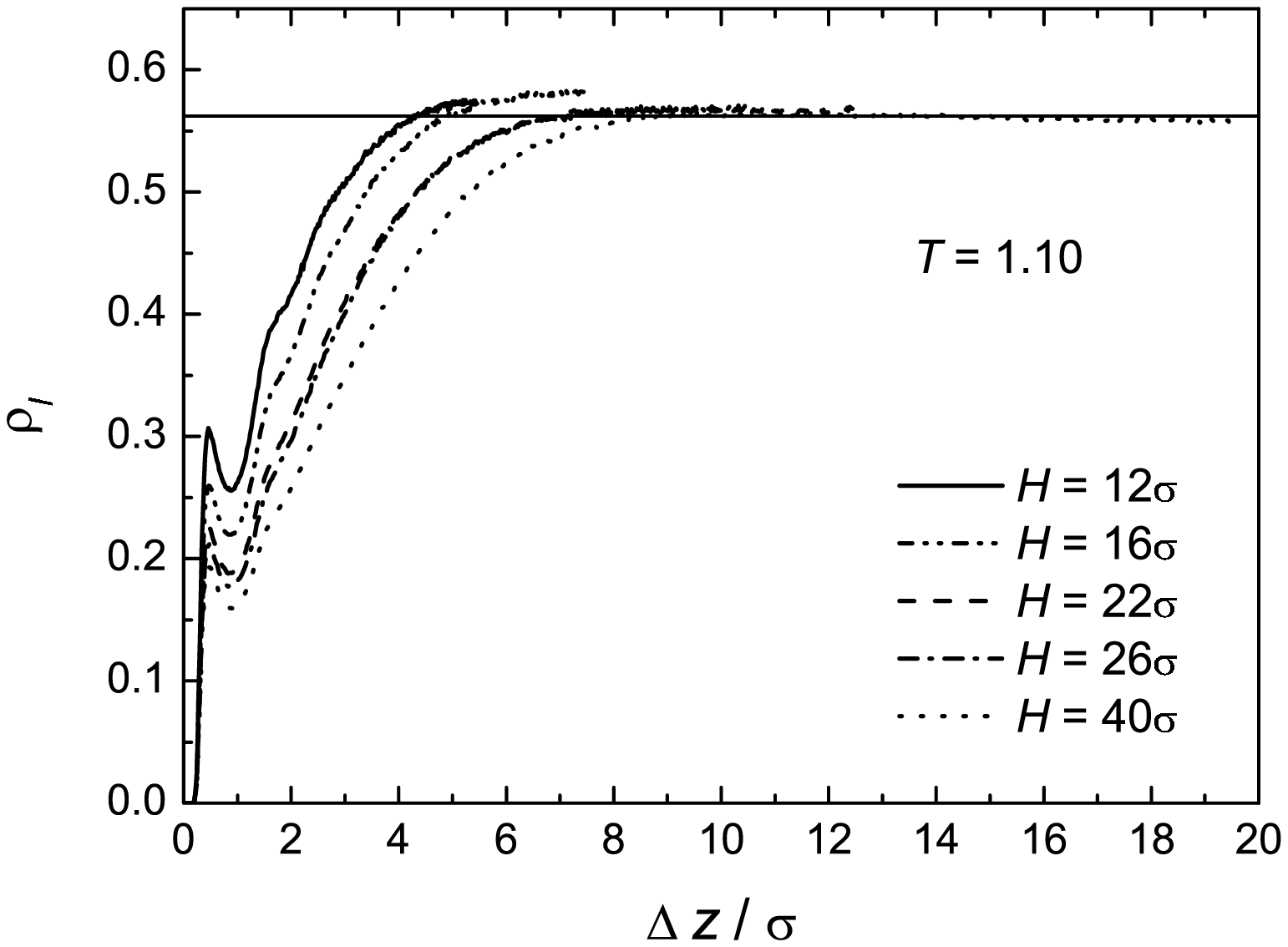}
\caption{Liquid density profiles $\rho_l$($\Delta${\it z},$\tau$) at {\it T} =
  1.10 in pores of various width with identical fluid-surface interaction potential, described by equation (\ref{poten1}). The liquid density of the bulk fluid is shown by a horizontal line.}
\end{center}
\end{figure}
\par 
The drying layer, clearly seen in a large pore with {\it H} = 40 $\sigma$, disappears in a small pore with {\it H} = 12 $\sigma$ and the same fluid-surface potential. This agrees with the theoretical expectation of strong sensitivity of the thickness of the drying layer to the chemical potential, which is shifted due to the confinement in the pore \cite{Dietrichrev}. One may expect, that the thickness of a drying layer {\it L$_0$} should continuously increase with increasing pore size and achieve some saturation value {\it L$_{0,inf}$} in a semi-infinite system. To check this expectation we have simulated liquid-vapor coexistence of the LJ fluid in several pores of various sizes with the same fluid-surface interaction potential at {\it T}  = 1.10. The obtained liquid density profiles shown in figure 10 indicate, that the density depletion becomes more pronounced in larger pores. The fits of $\rho_l$($\Delta${\it z},$\tau$) with equation (\ref{tanh}) allow an estimation of the parameters {\it L$_0$} and $\xi_-$ in all studied pores, and their dependences on the pore size are presented in figure 11. Note, that at this temperature, {\it L$_0$} exceeds 1$\sigma$ and the drying layer is indeed detectable in all pores except the smallest one with {\it H} = 12 $\sigma$. As the shift $\Delta\mu$ of the chemical potential of the phase transition in a pore relatively to a semi-infinite system is proportional to 1/{\it H} \cite{Kelvin}, it is reasonable to consider the dependence of {\it L$_0$} and $\xi_-$ on 1/{\it H}. Both parameters show essentially a linear dependence (see figure 11). Extrapolation of these linear dependences to semi-infinite system (1/{\it H} $\rightarrow$ 0) gives the values: {\it L$_{0,inf}$} = (2.67 $\pm$ 0.15) $\sigma$ and $\xi_{-,inf}$ = (1.65 $\pm$ 0.04) $\sigma$. Assuming that a drying layer effectively decreases  the pore size to {\it H} - 2 {\it L$_0$}, the dependences shown in figure 11 become non-linear and their extrapolation to semi-infinite systems by second-order polynomial fit gives {\it L$_{0,inf}$} = (2.79 $\pm$ 0.47) $\sigma$ and $\xi_{-,inf}$ = (1.73 $\pm$ 0.09) $\sigma$.  
 The values of $\xi_-$, extrapolated to a semi-infinite system, correspond to an amplitude of the correlation length $\xi_0$ $\approx$ 0.32-0.33 $\sigma$. This value is about 15 $\%$ higher the value $\xi_0$ = 0.28 $\sigma$ obtained in the pore with {\it H} = 40 $\sigma$. 
\begin{figure}
\begin{center} 
\includegraphics [width=8cm]{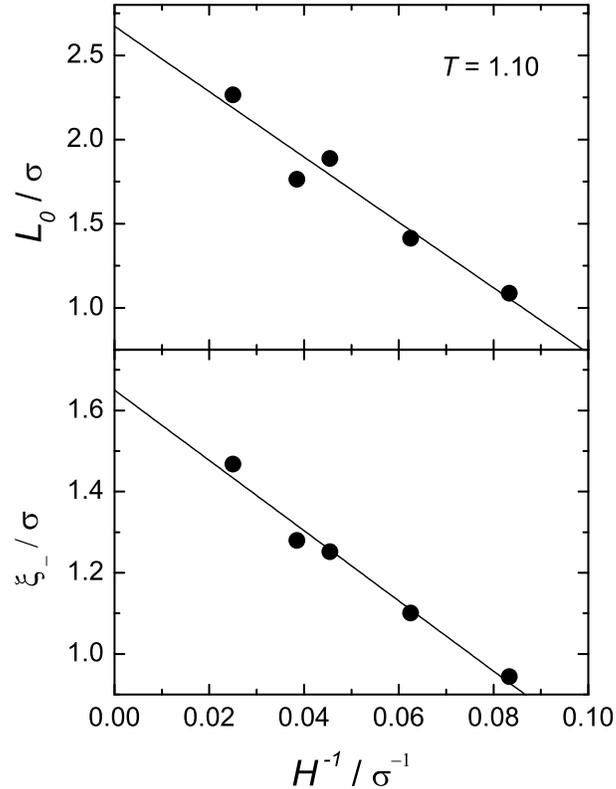}
\caption{Dependence of the thickness of the drying layer {\it L$_0$} and the correlation length $\xi_-$, obtained from fits of the liquid density profile $\rho_l$($\Delta$z,$\tau$) with equation (\ref{tanh}), on the inverse pore width {1/\it H} (circles). Linear fits of the data are shown by lines.}
\end{center}
\end{figure}
\begin{figure}
\begin{center} 
\includegraphics [width=8cm]{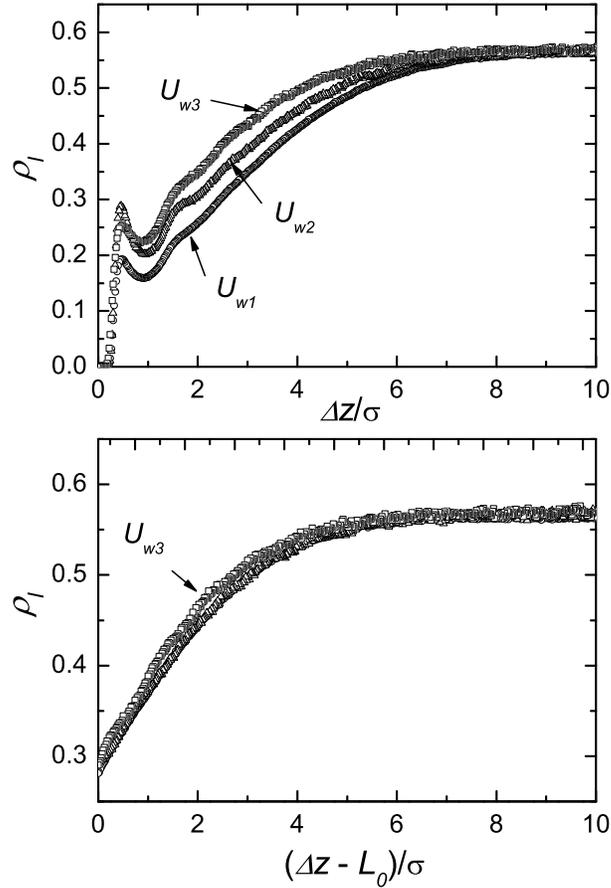}
\caption{Upper panel: liquid density profiles $\rho_l$($\Delta$ z,$\tau$) at {\it T} = 1.10
(circles) in pores of width {\it H} = 40 $\sigma$ with
{\it{U$_{w}$(z)}}(circles), {\it{U$_{w1}$(z)}} (triangles) and
{\it{U$_{w2}$(z)}} (squares) fluid-wall interaction potentials. Lower panel: the same profiles shown as a distance to the inflection points (see text). }
\end{center}
\end{figure}
\par 
Finally, we have tested the general theoretical expectation, that the formation of a drying layer could be suppressed by a strengthening fluid-surface interaction or by extending its attractive range. For this purpose we have simulated liquid-vapor coexistence and density profiles of the coexisting phases in the large pores ({\it H} = 40 $\sigma$) at {\it T} = 1.05, 1.10 and 1.15 with stronger interaction potential {\it{U$_{w2}$(z)}} and with the slower decaying potential {\it{U$_{w3}$(z)}} (see section Methods for more details). The obtained density profiles at {\it T} = 1.10 are compared in figure 12. Evidently, the drying layer is strongly influenced by the interaction potential. Namely, strengthening of the fluid-surface potential by about 33 $\%$ causes a shrinking of a drying layer at {\it T} = 1.10 from 2.26 $\sigma$ to 1.76 $\sigma$, i.e. by about 20 $\%$. At {\it T} = 1.05 the effective thicknes of the drying layer is suppressed to about 1 $\sigma$. An extension of the attractive range of the fluid-surface potential from $\sim$ {\it r}$^{-4}$ to $\sim$ {\it r}$^{-3}$ has an even stronger effect: the drying layer almost dissappears ({\it L$_0$} $<$ 1.4 $\sigma$ at {\it T} = 1.10). 
\par
Starting from the inflection point, the shape of the liquid density profiles remains highly universal for all considered fluid-surface potentials, as it is determined mainly by the bulk correlation length (see lower panel in figure 12). Note, that the details of the long-range attractive tail of a fluid-surface potential can produce some minor deviations of $\rho_l$($\Delta$z,$\tau$) from the universal shape, described by equation (\ref{tanh}). An example of such deviations is also shown in the inset of figure 5.         
\begin{figure}
\begin{center} 
\includegraphics [width=8cm]{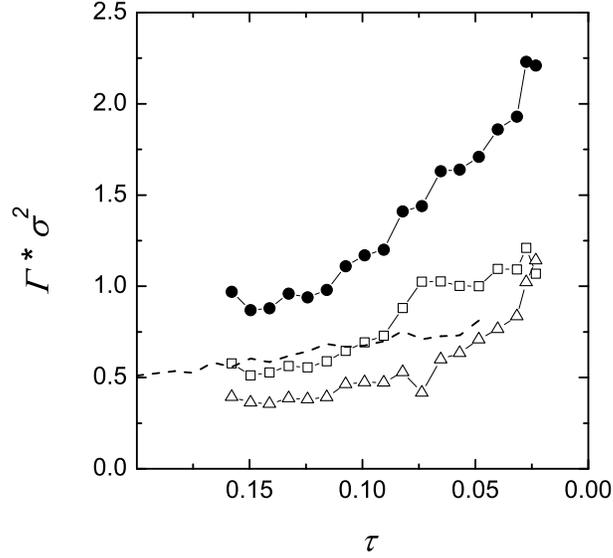}
\caption{Temperature dependence of the excess desorption of the liquid in the
  large pore: total excess desorption {\it $\Gamma_{tot}$} (solid circles) and
  its contributions due to drying layer {\it $\Gamma_{L}$}(squares) and due to increase of the
  correlation length  {\it $\Gamma_{\xi}$} (triangles). Total desorption in the
    small pore is shown by dashed line.}
\end{center}
\end{figure}
\par 
   Using the obtained density profiles of the liquid, we analyzed the surface excess desorption,
which describes a deficit of mass per unit surface area caused by depletion of the liquid
density near the surface. The total excess desorption {\it $\Gamma_{tot}$} contains two contributions: desorption due to the presence of a drying layer {\it $\Gamma_{L}$} and desorption due to the density depletion {\it $\Gamma_{\xi}$}, governed by the correlation length. We calculated these two contributions by numerical integration of $\rho_l$($\Delta$z,$\tau$) from the surface to the inflection point {\it $\Gamma_{L}$} and from the inflection point to the pore center {\it $\Gamma_{\xi}$}. The total excess desorption {\it $\Gamma_{tot}$} and its two contributions are shown in figure 13 as functions of the reduced temperature. For a semi-infinite system the excess desorption {\it $\Gamma_{\xi}$} can be simply calculated by integration of equation \ref{tanh} for the liquid-drying layer interface from the inflection point to infinity:
\begin{eqnarray} 
\label{gammaxi}
\Gamma_{\xi}(\tau) = \rho_{l,bulk}~\xi_-~ln2.
\end{eqnarray}
This part of the total desorption behaves like $\sim$ $\tau^{-\nu}$ at any
temperature and strongly diverges when approaching the critical point. Neglecting the density oscillations near the surface, the contribution {\it $\Gamma_{L}$} can be obtained in a similar way using integration of equation \ref{tanh} for the liquid-drying layer interface from the pore wall to the inflection point:
\begin{eqnarray} 
\label{gammaL}
\Gamma_{L}(\tau) = \rho_{l,bulk} \left[ L_0 + 2\xi_- ln \left( cosh(L_0/2\xi_-)\right) \right]/2
\end{eqnarray}
\par
We analyzed the ability of equations (\ref{gammaxi}) and (\ref{gammaL}) to describe the simulation results of the excess desorption. The excess desorption {\it $\Gamma_{\xi}$} depends on the bulk fluid properties only. Knowing the temperature behavior of the bulk liquid density $\rho_{l,bulk}$ along the coexistence curve \cite{lj1} and using the asymptotic equation (\ref{ksi}) for the bulk correlation length with amplitude $\xi_0$ = 0.28 $\sigma$, we can directly obtain the temperature dependence of {\it $\Gamma_{\xi}$} from equation (\ref{gammaxi}) (dashed line in figure 14). Good agreement of the simulated values of {\it $\Gamma_{\xi}$} with equation (\ref{gammaxi}) is observed in the whole studied temperature range. Deviations at low and high temperatures should be attributed to deviations of the correlation length from the asymptotic equation (\ref{ksi}) (see figure 6).  
\par
The excess desorption  {\it $\Gamma_L$}, described by equation (\ref{gammaL}), apart from the bulk parameters $\rho_{l,bulk}$ and $\xi_-$ contains also the parameter {\it L$_0$}, determined by the fluid-surface interaction. We obtain meaningful estimates of {\it L$_0$} for the temperatures, where the thickness of the drying layer {\it L$_0$} exceeds about 1 $\sigma$. In this range {\it L$_0$} follows equation \ref{layer} (see figure 6) and therefore the contribution {\it $\Gamma_L$} varies roughly as ln($\tau^{-1}$)({\it $\rho_{l,bulk}$)/2} , i.e. shows logarithmic divergence when approaching {\it T$_c$}. The excess desorption {\it $\Gamma_L$} described by equation \ref{gammaL}, and the total excess desorption {\it $\Gamma_{tot}$} = {\it $\Gamma_{\xi}$} + {\it $\Gamma_L$}, described by equations \ref{gammaxi} and \ref{gammaL}, are shown in figure 14 by dot-dashed and solid lines, respectively. Obviously, good agreement of the simulated values of {\it $\Gamma_L$} with equation (\ref{gammaL}) is observed in the range, where the temperature of {\it L$_0$} is known.
\begin{figure}
\begin{center} 
\includegraphics [width=8cm]{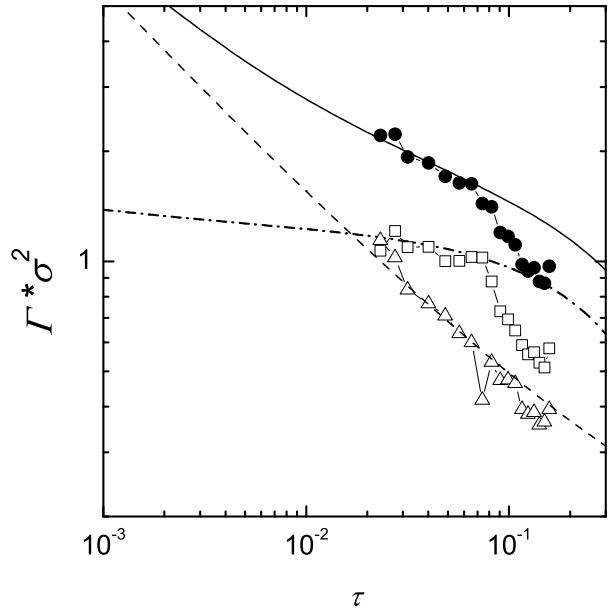}
\caption{Temperature dependence of the excess desorption of the liquid in
  double logarithmic scale: symbols are the same as in figure 14. Desorption
  {\it $\Gamma_{\xi}$} and {\it $\Gamma_L$}  described by equations
  \ref{gammaxi} and \ref{gammaL} and {\it $\Gamma_{tot}$} =  
{\it $\Gamma_{\xi}$} + {\it $\Gamma_L$} are
    shown by dashed, dashed-dot and solid lines, respectively.}
\end{center}
\end{figure}
\par
In a wide temperature range the contribution from the drying layer to the excess desorption is dominant (figure 13). However, the contributions {\it $\Gamma_L$} and {\it $\Gamma_{\xi}$} approach each other with increasing temperature and practically coincide at $\tau$ $\approx$ 0.02 (figure 14). The contribution {\it $\Gamma_{\xi}$} obviously should dominate {\it $\Gamma_L$} at higher temperatures, as the former diverges as $\tau^{-\nu}$, whereas the latter diverges logarithmically only.
\begin{figure}
\begin{center} 
\includegraphics [width=9cm]{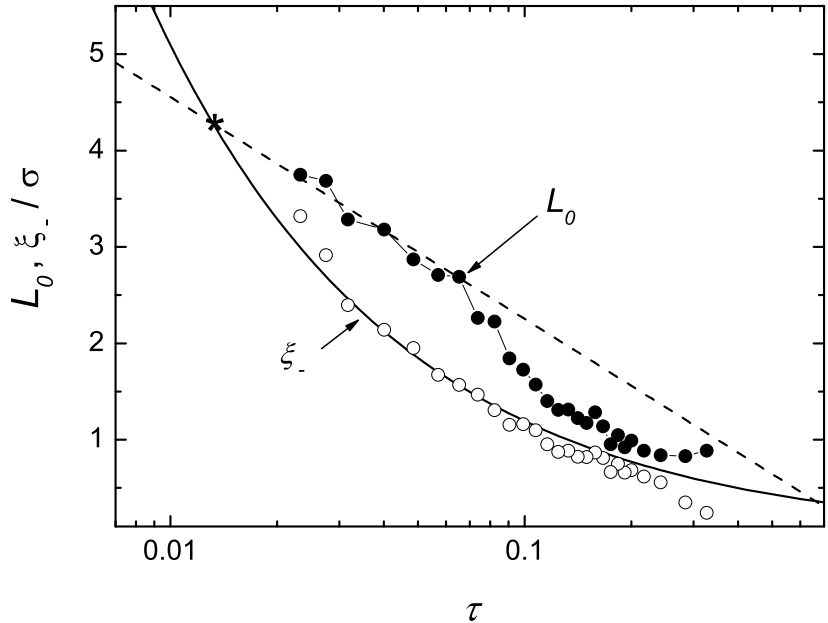}
\caption{Temperature dependence of the thickness of the drying layer
{\it L$_0$} (solid circles) and correlation length $\xi_-$ (open circles),
 obtained from fits
of the liquid density profile $\rho_l$($\Delta$ z,$\tau$), which extend from
the pore center to the distance 1 $\sigma$ from the surface.
The logarithmic law described by
equation (\ref{layer}) with {\it const} = -0.05 $\sigma$ is shown by dashed line. The power law 
(\ref{ksi}) with $\xi_0$ = 0.28 $\sigma$ is shown by solid
line. The temperature, where {\it L$_0$} $\approx$ $\xi_-$ is marked by a star.}
\end{center}
\end{figure}  
\section{Discussion}
We have studied the temperature evolution of the density profiles of a liquid along the liquid-vapor coexistence curve near a weakly attractive surface. To approach the fluid behavior in a semi-infinite system, the liquid-vapor coexistence curve of a LJ fluid was simulated in an extremely large slitlike pore with {\it H} = 40 $\sigma$. Additionally, at some temperatures we simulated the liquid-vapor coexistence in pores of various sizes. Strong depletion of the liquid density near a weakly-attractive pore wall is observed. The shape of the liquid density profiles evidences the formation of a drying (not a vapor) layer near the pore wall with increasing temperature. This drying layer can be detected at {\it T} $>$ 1.0 ( $\tau<$ 0.15), when the distance between the inflection point of the density profile and the pore wall exceeds at least one molecular diameter. We propose to describe the liquid density profiles near weakly attractive surfaces by equation (\ref{tanh}), which assumes an interface between the liquid phase and the drying layer and allows estimation of the bulk correlation length $\xi_-$ and the thickness of the drying layer {\it L$_0$}.  
\par 
The fits of $\rho_l$($\Delta${\it z},$\tau$) with the interfacial-like equation (\ref{tanh}) and with the exponential equation (\ref{exp}) which describes the decay to the bulk density far from the surface, give similar values of the amplitude of the correlation length $\xi_0$ $\approx$ 0.28 $\sigma$. This evidences a self-consistency of the proposed treatment of the liquid density profile and also indicates a negligible effect of the capillary waves on the studied interface between the liquid and the drying layer. This conclusion corroborates general theoretical arguments, concerning the influence of a long-range fluid-surface potential on the interface between vapor and wetting layer \cite{Dietrichrev}. 
\par 
The thickness of the drying layer {\it L$_0$} follows a logarithmic temperature dependence (\ref{layer}), starting from {\it T} $\approx$ 1.1 (i.e., at $\tau$ $<$ 0.07). Such a logarithmic divergence with approaching wetting (drying) temperature is expected for critical wetting (drying) \cite{Dietrichrev}. This result is consistent with the existence of a second-order drying transition at the bulk critical temperature for long-range fluid-surface interactions \cite{Ebner} and with experimental evidences for the absence of a drying transition at subcritical temperatures even in a system with extremely weak fluid-surface interaction \cite{Chan,NeCs}, similar to the one, used in our studies. Additionally, we have shown how the thickness {\it L$_0$} of the drying layer can be strongly suppressed by strengthening of the fluid-surface interaction, extension of its attractive range and by confinement. 
\par   
 In the temperature range of our simulation studies, the correlation length $\xi_-$ does not exceed the thickness of the drying layer {\it L$_0$} (see Table 1 and figure 15). Since the divergence of the correlation length $\xi_-$ is much stronger than the logarithmic divergence of {\it L$_0$}, when approaching the critical point $\xi_-$ should exceed {\it L$_0$}.
At the characteristic temperature $\tau \approx$ 0.01, shown by a star in figure 8, $\xi_-$ $\approx$ {\it L$_0$}. Above this temperature the interface between liquid and drying layer approaches the surface in terms of correlation length. The effective "thickness" of the total region of density depletion near the surface should diverge asymptotically as the correlation length. This corroborates the theoretical expectations for the thickness of the wetting (drying) layer, when the critical wetting (drying) transition occurs at the bulk critical temperature \cite{Indekeu}. Experiments with fluid mixtures indicate, that the partial wetting layer decays exponentially with distance from the surface and its effective "thickness" is proportional to the bulk correlation length $\xi_-$ \cite{Bonnrev,Christ2}. 
\par 
We found that the total excess desorption due to the depletion of the liquid density near a weakly attractive surface should diverge as the bulk correlation length $\xi_-$  $\sim$ $\tau^{-\nu}$ = $\tau^{-0.63}$ when approaching the critical temperature. This strong divergence, described by equation (\ref{gammaxi}), originates from that part of the liquid density profile which extends from the bulk liquid to the inflection point of the liquid-drying layer interface. Slight variations of the fluid-surface potential do not affect noticeably this part of the profiles (see figure 12, lower panel) and, accordingly, the value of {\it $\Gamma_{\xi}$}, whereas the thickness of the drying layer is highly sensitive to the details of the fluid-surface potential (figure 12, upper panel). 
\par
The observed divergence of the excess desorption is stronger than {\it $\Gamma$} $\sim$ $\tau^{\beta -\nu}$ = $\tau^{-0.305}$, expected from the theory of critical adsorption \cite{adsorption}. We do not see the possibility of a crossover of {\it $\Gamma$} to the power law of critical adsorption with further approaching the critical temperature, as the inflection point of the liquid-drying layer interface  asymptotically approaches the value $\rho_c$/2. Such a crossover could be possible only if the drying layer transforms into a vapor layer and, accordingly, the density at the inflection point of the liquid-vapor interface approaches $\rho_c$ with increasing temperature. However, this scenario is forbidden for systems with long-range fluid-surface interaction, where vapor layer do not appears below the critical temperature \cite{Ebner,Indekeu,Bruin}. Note, that recent experimental studies of the adsorption in binary mixtures indicate, that at subcritical temperatures {\it $\Gamma$} diverges stronger than $\sim$ $\tau^{\beta -\nu}$ \cite{Christ2,Fin2}. Such behavior may have the same origin, as that observed in our computer simulations. 

\end{document}